# Spin-orbit quantum impurity and quantization in a topological magnet

Jia-Xin Yin[1*†], Nana Shumiya[1*], Yuxiao Jiang[1*], Huibin Zhou[2*], Gennevieve Macam[3*], Hano Omar Mohammad Sura[4*], Songtian S. Zhang[1], Zijia Cheng[1], Zurab Guguchia[1,5], Yangmu Li[6], Qi Wang[7], Maksim Litskevich[1], Ilya Belopolski[1], Xian Yang[1], Tyler A. Cochran[1], Guoqing Chang[1], Qi Zhang[1], Zhi-Quan Huang[3], Feng-Chuan Chuang[3], Hsin Lin[8], Hechang Lei[7], Brian M. Andersen[4], Ziqiang Wang[9], Shuang Jia[2], and M. Zahid Hasan[†1,10]

[1]Laboratory for Topological Quantum Matter and Advanced Spectroscopy (B7), Department of Physics, Princeton University, Princeton, New Jersey 08544, USA.

[2]International Center for Quantum Materials and School of Physics, Peking University 100193, Beijing, China.

[3]Department of Physics, National Sun Yat-Sen University, Kaohsiung 80424, Taiwan.

[4]Niels Bohr Institute, University of Copenhagen, Universitetsparken 5, DK-2100 Copenhagen, Denmark.
[5]Laboratory for Muon Spin Spectroscopy, Paul Scherrer Institute, CH-5232 Villigen PSI, Switzerland.

[6]Condensed Matter Physics and Materials Science Division, Brookhaven National Laboratory, Upton, New York 11973, USA.

[7]Department of Physics and Beijing Key Laboratory of Opto-electronic Functional Materials & Micro-nano Devices, Renmin University of China, Beijing 100872, China.

[8]Institute of Physics, Academia Sinica, Taipei 11529, Taiwan.
[9]Department of Physics, Boston College, Chestnut Hill 02467, MA, USA.

[10]Materials Sciences Division, Lawrence Berkeley National Laboratory, Berkeley, CA 94720, USA.

†Corresponding authors, E-mail jiaxiny@princeton.edu; mzhasan@princeton.edu

*These authors contributed equally to this work.

## Abstract

**Quantum states induced by single-atomic impurities are at the frontier of physics and material science. While such states have been reported in high-temperature superconductors and dilute magnetic semiconductors, they are unexplored in topological magnets which can feature spin-orbit tunability. Here we use spin-polarized scanning tunneling microscopy/spectroscopy (STM/S) to study the engineered quantum impurity in a topological magnet $Co_3Sn_2S_2$. We find that each substituted In impurity introduces a striking localized bound state. Our systematic magnetization-polarized probe reveals that this bound state is spin-down polarized, in lock with a negative orbital magnetization. Moreover, the magnetic bound states of neighboring impurities interact to form quantized orbitals, exhibiting an intriguing spin-orbit splitting, analogous to the splitting of the topological fermion line. Our work collectively demonstrates the strong spin-orbit effect of the single-atomic impurity at the quantum level, suggesting that a nonmagnetic impurity can introduce spin-orbit coupled magnetic resonance in topological magnets.**

## Introduction

Understanding the single-atomic impurity state in a quantum material is a fundamental problem with widespread implications in physics and technology[1-8]. For instance, the Zn impurity state in a high-temperature superconductor uncovers the Cooper pairing symmetry[3], the Mn impurity state in a semiconductor elucidates the ferromagnetic coupling[4], and the interstitial Fe impurity in a superconductor with topological surface states creates Majorana-like state[5]. Besides being a local probe of the quantum materials, the impurity state with discrete or (magnetic) bistable quantum levels is valuable for the quantum technology[6-8]. Most known single-atomic impurity states are, however, either from the spin or orbital channel, limiting their tunability at the quantum level. Recently, spin-orbit coupled magnets have emerged as a new class of quantum materials suitable for microscopic research[9-19]. In particular, we notice that the In doped $Co_3Sn_2S_2$ exhibits strongly altered bulk magnetic and transport properties, including reductions of magnetism, suppressions of metallicity, and variations of anomalous Hall conductivity[20-22]. These effects imply a striking, yet not understood quantum state associated with each nonmagnetic In impurity in this topological magnet. Therefore, a single crystal of $Co_3Sn_2S_2$ containing a dilute concentration of In impurities is considered a natural and promising quantum material for experiments on atomic impurity state with spin-orbit tunability. Here we report our spin-polarized STM/S study of 1% In doped $Co_3Sn_2S_2$, which uncovers a spin-orbit quantum impurity state.

## Results

**Engineered atomic impurity.** $Co_3Sn_2S_2$ has a layered crystal structure and a ferromagnetic ground state (Curie temperature, $T_C$ = 170K) with the *c*-axis magnetization arising from the Co kagome lattice. Cleaving preferentially breaks its S-Sn bond, which leads to the S and Sn terminated surfaces. Previous STM studies have dominantly observed two surfaces, one with largely vacancy defects and the other with adatom defects[11,13,14]. Several factors challenge their assignation, as the two surfaces have identical lattice symmetries, their interlayer distance is sub-Å in scale, and STM topographic image convolutes the spatial variation of the integrated local density of states and the geometrical corrugations[23]. Decisive experimental evidence for surface identification can be found by imaging the symmetry-dictated surface boundary and the layer-selective chemical dopants[23-25]. In previous work[11], we have determined that the vacancy surface is the Sn layer and the adatom surface is the S layer by imaging their surface boundary determined by the crystalline symmetry. We further conclude this assignation by doping the bulk $Co_3Sn_2S_2$ single crystals with 1% In impurities and imaging the layer-selective In-dopants. The In impurities preferentially replace the Sn atoms according to previous experimental and theoretical studies[20,21], as well as our recent systematic single-crystal growth[22]. Indeed, on the vacancy surface identified[11] to be Sn, we observe dilute substitutional atoms with consistent concentration (Fig. 1**a**), suggesting these impurities to be In atoms.

**Spatial feature of single impurity resonance.** By performing an extensive study on the electronic properties of the In impurities, we find that each impurity repeatedly features a sharp state at the negative energy as shown in Fig. 1**b**. First-principles calculations show that each In impurity is almost nonmagnetic but introduces a strong resonance (Fig. 1**c**), similar to the experimental data. The calculations further reveal that the impurity resonance arises from a spin-down state (opposite to the bulk magnetization direction) and resides in the spin-down bandgap. Thus, the magnetic resonance state likely arises from the local impurity perturbation of the spin-polarized band structure. As the low-energy band structure is dominated by Co 3$d$ orbitals, the resonant impurity state of the In atom also implies that there a strong local impact on the Co kagome lattice in real-space. To explore the detailed real-space feature, we probe the local electronic structure for an isolated In impurity, as shown in Fig. 1**d**. The corresponding d$I$/d$V$ map at the impurity resonance energy in Fig. 1**e** shows a localized pattern bound to the impurity site (Fig. 1**e**). The bound state couples with three nearby Co atoms in the underlying kagome lattice, as illustrated in Fig. 1**f**. This is consistent with the first-principles calculation that the nearby magnetic Co atoms also feature such resonance state (inset of Fig. 1**c**), supporting the Co-In coupling (See Supplementary Note 1). Figure 1**g** shows the representative d$I$/d$V$ curves measured with increasing distance from the impurity, demonstrating the bound state decaying in intensity without detectable energy dispersion or splitting. An exponential fit to the decay yields a characteristic length scale of 2.8Å (inset of Fig. 1**g**). Therefore, these systematic characterizations reveal that the nonmagnetic In impurity couples with the underlying magnetic kagome lattice to introduce a striking localized bound state.

**Magnetic nature of single impurity resonance.** To probe the magnetic nature of the impurity bound state, we perform tunneling experiments with a spin-polarized Ni tip under weak magnetic fields[26-29]. The bulk crystal has a coercive field $B_C \sim 0.3$T, and Ni tip is a soft magnet with a $B_C \ll$ 0.1T that can be easily flipped by reversing the magnetic field[29]. We measure the tunneling signal of the impurity state while sequentially applying fields along the $c$-axis of +0.5T, +0.1T, -0.1T, -0.5T, -0.1T, and +0.1T to systematically flip the magnetization of the tip and sample (Fig. 2**a**). This sequence allows us to perform spin-polarized measurements of the impurity. The +0.5T field polarizes both the sample and tip, aligning the spin of the tip and anti-aligning the spin of the impurity state, due to the spin-down nature of the impurity state. A +0.1T field does not change the polarization of either the tip or impurity state. Flipping the field to -0.1T also flips the spin of the tip, leaving the spin of the impurity state unchanged (down). Here, with both tip and impurity state spins aligned down, we observe an intensity increment of the tunneling signal. Next, we further decrease the field to -0.5T, which flips up the spin of the impurity state with a corresponding reduction of the tunneling signal. Lastly, by sequentially changing the field to -0.1T and +0.1T, we flip the spin of the tip (down) and again observe an increase in the intensity. Our progressive field manipulation strongly supports that impurity state features spin-down polarization tied to the bistable magnetic bulk, consistent with the first-principles calculation.

To further determine the effective moment of this magnetic polarized state, we probe the state by applying a strong external magnetic field ($|\mathbf{B}| \gg B_C$) along the $c$-axis with a nonmagnetic tip. Under the field, a spin-up/spin-down band hosting an intrinsic magnetic moment of +1/-1 Bohr magneton will exhibit a Zeeman shift to lower/higher energies in a rate of 0.058meVT$^{-1}$. Moreover, when the magnetism of the system is polarized with an applied field, the spin-polarized state will always shift to the same energy direction regardless of the relative field orientation[11] (top inset schematic in Fig. 2**b**), which was also experimentally observed for the 8T and -8T data (Fig. 2**b**). The positive energy shift indicates the state has a negative magnetic moment, calculated to be -5$\mu_B$ (or a Landé $g$ factor of 10) based on a shift rate of 0.275meVT$^{-1}$ (right inset of Fig. 2**b**). This large value is beyond the spin Zeeman effects (~-1$\mu_B$) and indicates the additional negatively polarized orbital magnetization. The anomalous Zeeman effect with an unusual moment or *g* factor has been observed in the electronic bands of kagome magnets[9,11,17], which is often linked to the Berry phase physics associated with magnetism and spin-orbit coupling[9,11,30,31]. Since the In impurity couples with the Co atoms in the kagome lattice, the higher orbital angular momentum of the hybridized In-Co orbital can contribute to the large effective moment. We note that the effective moment of -$5\mu_B$ represents the diamagnetic response induced by the In impurity, and should not be thought as the local magnetic moment of the impurity atom.

**Interacting impurity states induced quantized orbitals.** Having characterized magnetic resonance state from the isolated impurity, we further probe the coupled impurity states to understand how they interact with each other through extensive imaging and spectroscopy investigation with a nonmagnetic tip. In Fig. 3**a**, we present the evolution of the impurity bound state with increasing perturbation strength caused by a second nearby impurity. We find that with decreasing spatial separation, the bound state progressively decreases in intensity and finally splits into two sub-peaks. Figure. 3**b** further compares three cases with one isolated impurity, two neighboring impurities, and three neighboring impurities, respectively. We find the quantized number of split impurity states matches with the coupled impurity number, highlighting their atomic-scale quantum-level coupling. Differential conductance maps at these corresponding splitting energies demonstrate their distinct orbital hybridizations (Fig. 3**c**). For two neighboring impurities, the d*I*/d*V* maps show a bonding (σ) and antibonding (σ*) orbital formation[4], consistent with the quantum coupling of doubly degenerate states. For three impurities, the dI/dV maps show the formation of one bonding (σ) and two antibonding ($\sigma_1$*, $\sigma_2$*) orbitals, an unusual situation in which we discuss below.

**Discussion**

Geometrically, the three neighboring impurities have $C_{3V}$ symmetry, which would form a doubly degenerate orbital state[7] σ* protected by the mirror symmetry. The mirror symmetry operation, however, would transfer spin-up to spin-down. On the other hand, the magnetic resonances induced by different nonmagnetic In impurities are expected to be of the same spin-polarization direction, which is locked by the ferromagnetic ordering of the underlying magnetic Co kagome lattice perturbed by these impurities (See Supplementary Note 2). Therefore, the combination of the same spin-polarization direction of the degenerate impurity states and the atomic spin-orbit coupling naturally breaks the mirror symmetry, leading to the splitting of σ* (see "Methods" for theoretical modeling). Such splitting is analogous to the splitting of the bulk topological nodal line or magnetic Weyl fermion line[32,33], which is protected by the crystalline mirror symmetry.

In conclusion, we report the first STM/S studies of the nonmagnetic impurity behavior in a topological magnet. Associated with the atomic nonmagnetic impurities, we find an intense spin-orbit polarized bound state with an unusual magnetic moment and quantized energy splitting under impurity-impurity interaction. The striking spin-orbit quantum states revealed here can advance the understanding of the magnetic and transport behaviors of doped topological magnets, and call for new perspectives on the interplay between magnetism and spin-orbit coupling at the atomic scale. The discrete quantum level of interacting impurities resembles that of a quantum dot, which is critical to nanophotonics and quantum information processing. With the single-atom precision, the atomic quantum dot has a high level of digital fidelity[7]. The interacting spin-orbit polarized quantum impurity involves multiple degrees of freedom, including charge, spin, and orbital, the quantum control of which in the magnetic bistable platform can provide a useful guideline for the development of spin-orbit entangled quantum technology.

**Figure captions**

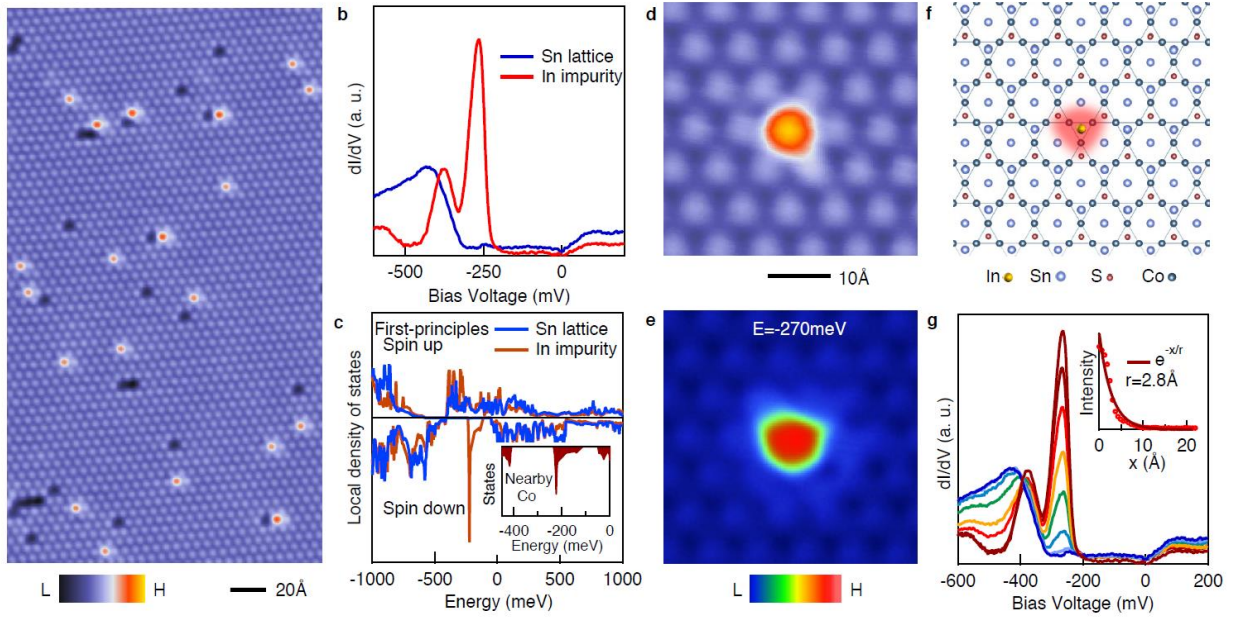

**Figure 1. Engineered atomic impurity state in a topological magnet. a** Atomically-resolved topographic image of Sn layer of 1% In doped $Co_3Sn_2S_2$. **b** Differential conductance spectrums taken on the Sn lattice (blue) and at the In impurity (red), respectively. **c** First-principles calculation[13] of the spin-resolved local density of states of the Sn lattice (blue) and an In impurity (red), which shows a magnetic impurity resonance. The inset shows the spin-down states of the Co atom closest to the In impurity. **d** Topographic image of an isolated impurity. **e** Corresponding differential conductance map taken at E = -270meV (resonance energy). **f** Correlation between the atomic structure and the pattern in the differential conductance map. **g** Differential conductance spectra taken across the surface with spatial variation from the center of the In impurity (dark red) to far away (blue). The inset shows the exponential fit to the spatial decay of the impurity resonance.

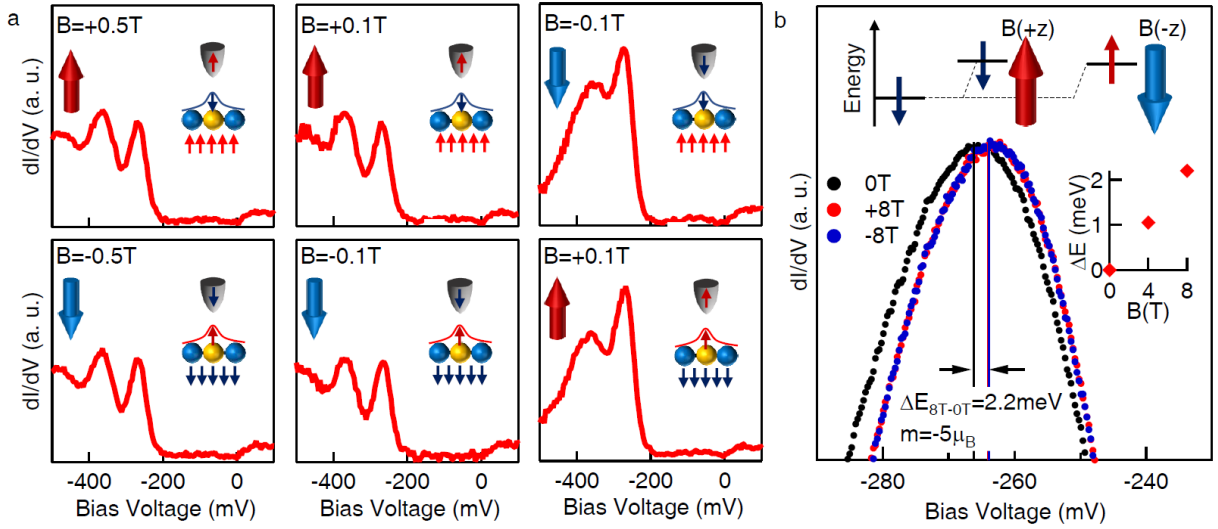

**Figure 2. Magnetic nature of the impurity bound state. a** Dependence of the impurity state with Ni tip under a weak magnetic field. We apply +0.5T, +0.1T, -0.1T, -0.5T, -0.1T, +0.1T fields to systematically flip the magnetization of the tip and sample. The inset schematics illustrate the respective spins of the tip and the impurity state that is anti-aligned with bulk magnetization direction. **b** Dependence of the impurity state with a strong magnetic field. Under both +8T and -8T, the peak exhibits a magnetization-polarized Zeeman energy shift of 2.2meV, which amounts to an effective moment of -5$\mu_B$. The inset data shows the energy shift for different magnetic field magnitudes. Inset schematic illustrates the magnetization-polarized Zeeman effect. The applied field aligns the spins of the impurity state in the same orientation, hence +z and -z orientation fields lead the energy to shift in the same direction.

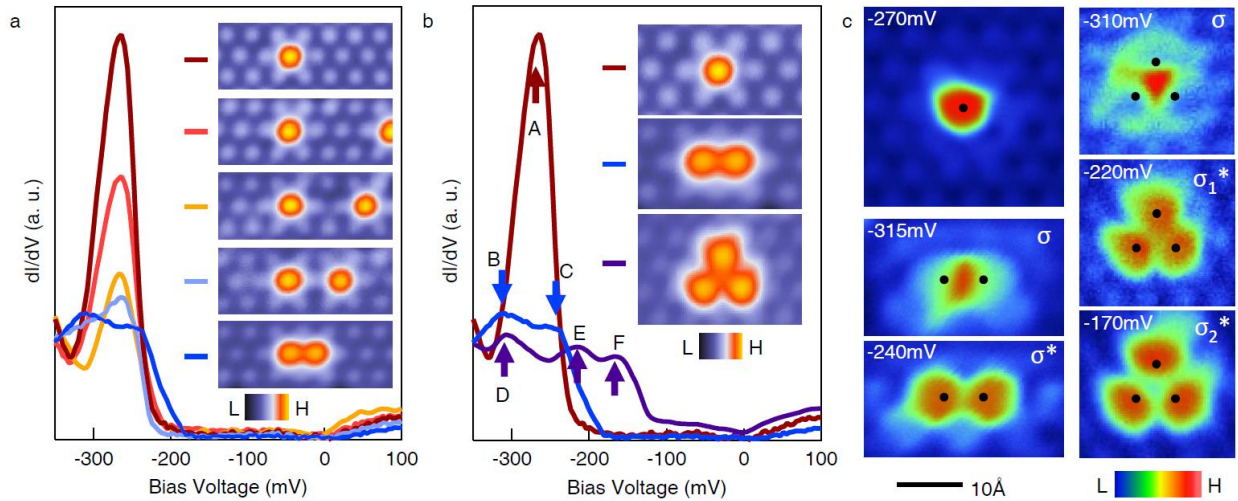

**Figure 3. Interacting impurity states induced quantized orbitals. a** Differential conductance spectra taken at the central impurity with perturbations of varying strengths from a second impurity. Inset: respective topographic images for impurity configurations. Note the images are acquired by finding different surface locations. **b** Local impurity state with coupling to different numbers of impurities. The arrows highlight the quantized splitting with additional interacting impurity numbers. Note the images are acquired by finding different surface locations. **c** dI/dV maps at the respective bound state energies in **b**. dI/dV maps taken at $V$ = -270mV for a single impurity; -315mV (bonding state σ) and -240mV (antibonding state σ) for a double impurity, respectively; -310mV (bonding state σ), -220mV (antibonding state $σ_1^*$) and -170mV (antibonding state $σ_2^*$) for a triple impurity, respectively. The black dots mark the center of impurities.

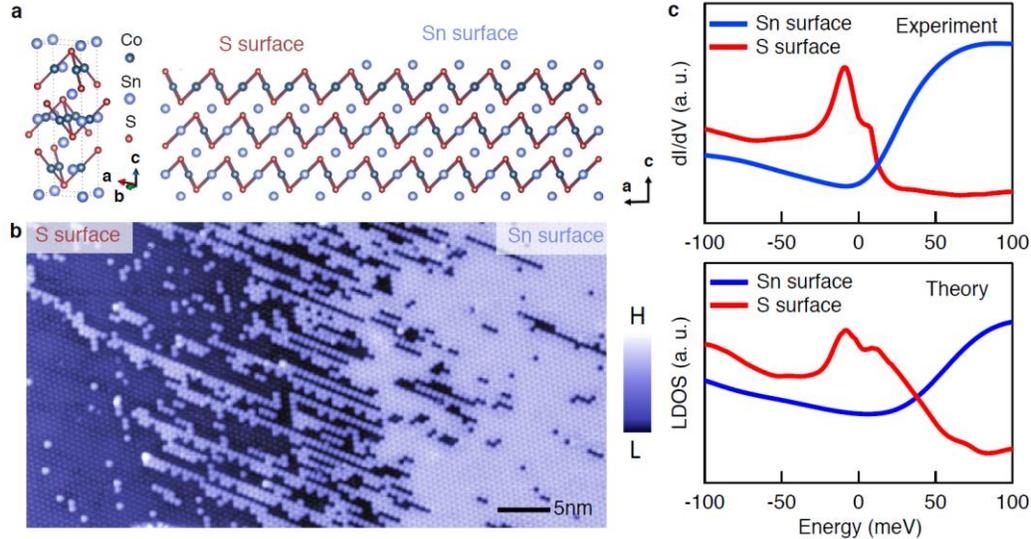

**Figure. 4 Extended evidence for surface identification in $Co_3Sn_2S_2$. a** Crystal structure of $Co_3Sn_2S_2$ (left) and the cleaving surface illustration (right). **b** Atomically resolved topographic image of the boundary between S surface and Sn surface. The S surface smoothly evolves into the Sn surface with increasing coverage of Sn adatom. **c** Comparison between spatially averaged surface dependent dI/dV curves with the first-principles calculations.

## Methods

**STM/S measurement.** Single crystals[22] of $Co_3S_{2-x}In_xSn_2$ up to 1.5mm × 1.5mm × 0.3mm were used in this study. Samples were cleaved mechanically in situ at 77K in ultra-high vacuum conditions, and then immediately inserted into the STM head, already at He4 base at 4.2K. The magnetic field was applied under zero-field cooling, after which we carefully approached the tip to locate the same atomic-scale area for tunneling spectroscopy[11,13]. Tunneling conductance spectra were obtained using standard lock-in amplifier techniques with a root mean square oscillation voltage of 0.2meV-5meV and a lock-in frequency of 973Hz. The topographic images were taken with tunneling junction set up: $V$ = -300~-500mV, $I$ = 100pA, the conductance maps were taken with tunneling junction set up: $V$ = -100~-500mV, $I$ = 200pA, and the tunneling spectra were taken with junction set up: $V$ = -600mV, $I$ = 300pA. Commercial STM Ir/Pt tip (nonmagnetic) and STM Ni tip (soft magnet) tips were used in this study. To study the impurity-impurity interaction, we checked the topographic images for over 3000 impurities to obtain the cases in Fig. 3 in the 1% In doped samples.

**Evidence for surface identification.** STM studies of $Co_3Sn_2S_2$ often encounter two kinds of surfaces, one dominated by adatom defects and the other dominated by vacancy defects. Here we discuss our evidence for their assignment[11] as S surface and Sn surface, respectively. Firstly, as illustrated in Fig. 4**a**, based on crystalline symmetry, when S and Sn surfaces meet at a step edge, the Sn surface will be just above the S surface. This experimental evidence is directly provided in Fig. 4**b**, where we observe the vacancy surface is above the adatom surface. Secondly, this identification provides a natural explanation for the origin of the surface defects. The Sn vacancy in the S surface and Sn adatom on the S surface are simultaneously created due to incomplete cleaving, as seen in the experiment (Fig. 4**b**). Thirdly, we show that the first-principles calculation of the surface dependent local density of states matches the experimental data (Fig. 4**c**). Lastly, we observe In dopant in the vacancy surface in the main paper, and In is known to substitute the Sn atom both experimentally and theoretically[20-22].

**First-Principles calculations.** First-principles calculations were performed in the density functional theory[34,35] framework as implemented in the Vienna Ab initio Simulation Package[36,37]. Generalized gradient approximation in Perdew−Burke−Ernzerhof functional[38] was applied to describe electron exchange-correlation interaction with the projector augmented wave potentials[39]. The supercell consists of a periodically repeating $2 \times 2$ -slab with a thickness of twice the bulk and a vacuum space of $\sim 14$ Å along the *z*-direction. The slab is cleaved to reveal the Sn-terminating surface, and one surface Sn atom is replaced with In to simulate dilute doping (~2.3%). The energy cutoff was set at 400 eV and the energies in self-consistent calculations were converged until $10^{-4}$ eV. The Brillouin zone was sampled using a $6 \times 6 \times 1$ Monkhorst-Pack[40] grid.

**Acknowledgments**

We thank Z. Song, T. Neupert, B. Lian and H. J. Gao for insightful discussions. Experimental and theoretical work at Princeton University was supported by the Gordon and Betty Moore Foundation (Grant No. GBMF4547 and GBMF9461/Hasan). Sample characterization was supported by the United States Department of Energy (US DOE) under the Basic Energy Sciences programme (Grant No. DOE/BES DE-FG-02-05ER46200). M.Z.H. acknowledges support from Lawrence Berkeley National Laboratory and the Miller Institute of Basic Research in Science at the University of California, Berkeley in the form of a Visiting Miller Professorship. This work benefited from partial lab infra-structure support under NSF-DMR-1507585. M. Z. H. also acknowledges visiting scientist support from IQIMat the California Institute of Technology. The work at Peking University was supported by the National Natural Science Foundation of China No. U1832214, No.11774007, the National Key R&D Program of China (2018YFA0305601) and the strategic Priority Research Program of Chinese Academy of Sciences (XDB28000000). The work at Renmin University was supported by the National Key R&D Program of China (Grants No. 2016YFA0300504 and 2018YFE0202600), the National Natural Science Foundation of China (No. 11774423,11822412), the Fundamental Research Funds for the Central Universities, and the Research Funds of Renmin University of China (RUC) (18XNLG14, 19XNLG17). Work at Boston College was supported by the U.S. Department of Energy, Basic Energy Sciences Grant No. DE-FG02-99ER45747. F.C.C. and H.L. acknowledge support from the National Center for Theoretical Sciences and the Ministry of Science and Technology of Taiwan under Grants No. MOST-107-2628-M-110-001-MY3 and MOST-109-2112-M-001-014-MY3. F.C.C. is also grateful to the National Center for High-Performance Computing for computer time and facilities. B.M.A. and H.O.M.S. acknowledge support from the Independent Research Fund Denmark grant number DFF 8021-00047B.

**Supplementary Note 1**

**Additional first-principles calculations.** To understand the local In-Co coupling, we calculate the charge densities of different systems as shown in Fig. 1. Figure 1a shows the charge density difference ($\Delta\rho$) between an In-doped $Co_3Sn_2S_2$ system ($\rho$(In-CSS)), an isolated In atom ($\rho$(In)), and a $Co_3Sn_2S_2$ system with a Sn vacancy ($\rho$(CSS-Sn vac)). The calculation shows that there is local charge coupling between the In atom and the Co kagome lattice atoms. Figure 1b shows another charge density difference ($\Delta\rho'$) between an In-doped $Co_3Sn_2S_2$ system ($\rho$(In-CSS)), and a $Co_3Sn_2S_2$ system ($\rho$(CSS)). This calculation shows that the additional charge variation induced by the In atom as a substitutional impurity also couples with the Co kagome lattice. Both calculations demonstrate the local In-Co charge coupling.

The local charge In-Co charge coupling also has an impact on the local magnetism as shown in Fig. 2. The local magnetic moment for each atom is calculated through the sum of all occupied spin-up and spin-down local density of states. The calculated magnetic structure of an In-doped $Co_3Sn_2S_2$ system (Fig. 2a) shows that the local moment mainly arises from the Co atoms in the kagome lattice, which reduces a little for the three Co atoms near the In impurity. Figure 2b shows more details of the calculated magnetic structure for the top In-Sn, S and Co layers. The moment of In impurity is -0.026$\mu_B$, which is an order of magnitude smaller than that of Co (~+0.3$\mu_B$), and of similar value as that of Sn lattice atom (~-0.02$\mu_B$ at far away positions or in the bulk). Since the magnetism mainly arises from the Co atom in the kagome lattice and no additional large local magnetic moment is directly introduced by the In impurity, it is more reasonable to treat the In atom as a nonmagnetic impurity in this system for the theoretical modeling.

We further perform a first-principles calculation of the triplet impurity case. The supercell consists of a periodically repeating 3×3-slab with a thickness of a bulk unit cell and a vacuum space of ∼14 Å along the $z$-direction. The slab is cleaved to reveal the Sn-terminating surface and three surface Sn atoms are replaced with In to simulate the In trimer impurity. The energy cutoff was set at 300 eV and the energies in self-consistent calculations were converged until $10^{-4}$ eV. The Brillouin zone was sampled using a Gamma-centered 3×3×1 Monkhorst-Pack grid. Without spin-orbit coupling, each In impurity exhibits two sharp spin-down polarized bound states in the local density of states calculation (Fig. 3a). The antibonding state σ* is much stronger than bond state σ, consistent with the symmetry analysis that σ* is doubly degenerate. With including spin-orbit coupling, σ* is splits into two peaks, and there are totally three impurity states (Fig. 3b), consistent with the experimental observation.

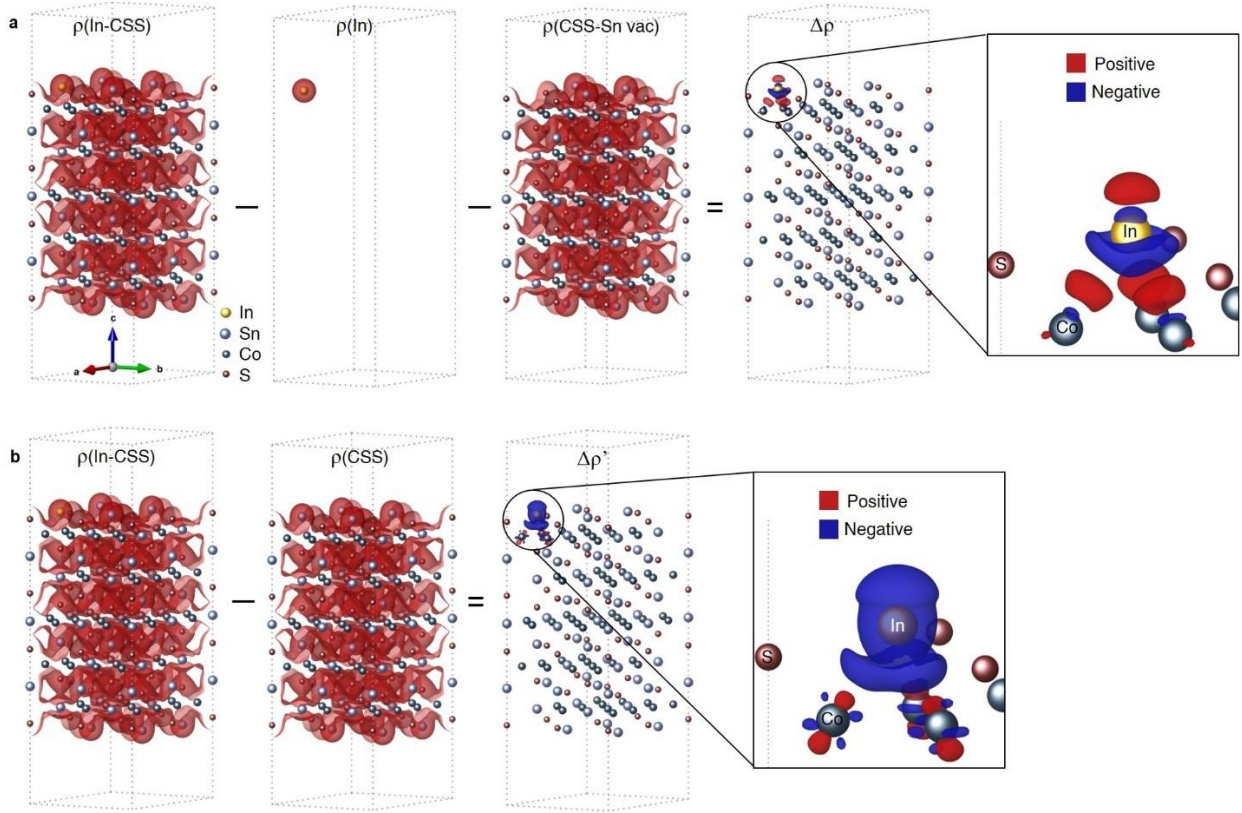

**Supplementary Figure 1: Local In-Co charge coupling. a** Charge density difference (Δρ) between an In-doped $Co_3Sn_2S_2$ system (ρ(In-CSS)), an isolated In atom (ρ(In)), and a $Co_3Sn_2S_2$ system with a Sn vacancy (ρ(CSS-Sn vac)). **b** Charge density difference (Δρ') between an In-doped $Co_3Sn_2S_2$ system (ρ(In-CSS)), and a $Co_3Sn_2S_2$ system (ρ(CSS)). The isosurface levels of the individual and differential systems are $8\times10^{-06}$ eÅ$^{-3}$ and $8\times10^{-07}$ eÅ$^{-3}$, respectively. The spin-orbit coupling is considered in this calculation.

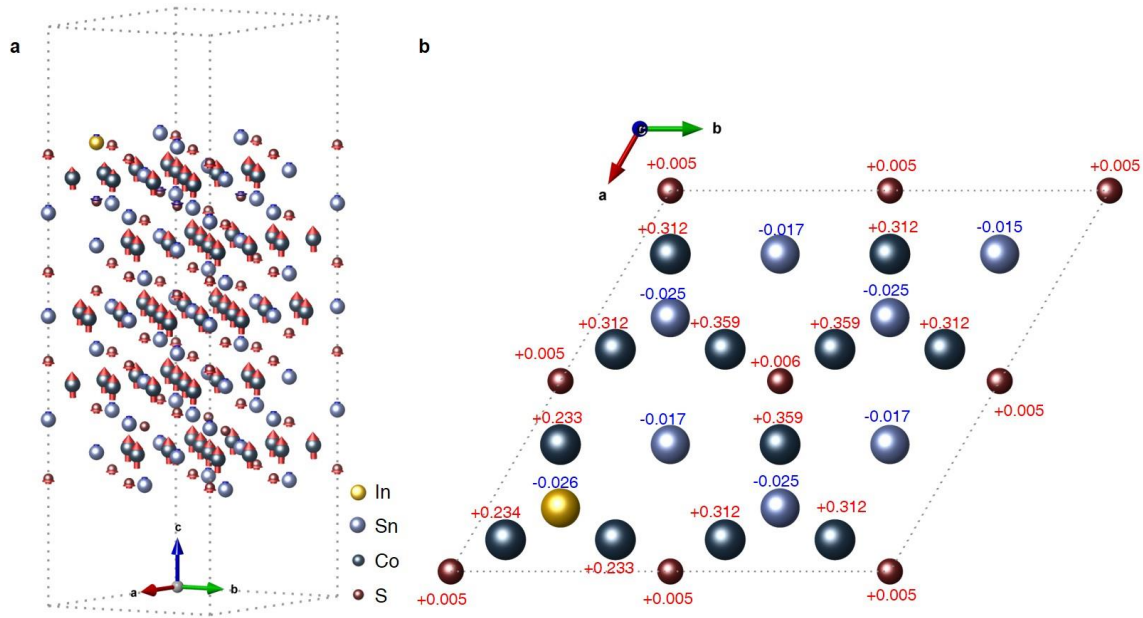

**Supplementary Figure 2: Impact of In impurity on local magnetism. a** Magnetic structure of the In-doped $Co_3Sn_2S_2$ system. The arrows in the atoms are proportional to the magnetic moment size. **b** Magnetic structure for the top In-Sn, S and Co layers. The magnetic moment is marked for each atom with the units of $\mu_B$. The spin-orbit coupling is considered in this calculation.

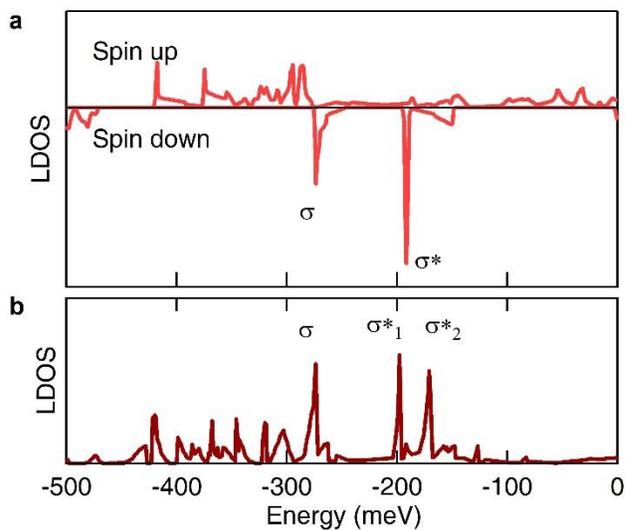

**Supplementary Figure 3: Impact of spin-orbit coupling on triple impurity state. a,** First-principles calculation of the spin-resolved local density of states at the In triple impurity without spin-orbit coupling. **b,** First-principles calculation of the local density of states at the In triple impurity with spin-orbit coupling.

## Supplementary Note 2

**A heuristic model for the magnetic impurity resonance** We show that nonmagnetic scattering in the spin-orbit coupled magnetic system can generate magnetic impurity resonance, by considering a nearest-neighbor hopping ($t$) model on the kagome lattice, with spin-orbital coupling ($\lambda$) and an overall Zeeman field ($B$) within the T-matrix approach. We assume that the relevant bands stem from a single orbital ($d_{z^2}$) at each Co-atom in the kagome lattice[13], and the In-Co coupling effectively introduces local potential scattering on the kagome lattice. The unit cell of the kagome lattice includes three Co-atoms. The lattice vectors are $\boldsymbol{a_1} = a(1,0), \boldsymbol{a_2} = a(1,\sqrt{3}), \boldsymbol{a_3} = \boldsymbol{a_2} - \boldsymbol{a_1}$. For the calculation of the impurity-free Green function, we identify the $\boldsymbol{k}$-vectors within the Brillouin zone for a certain lattice size. We assume that there are N=100 lattice points along the $\mathbf{a}_1$ and $\mathbf{a}_2$ primitive lattice vectors. The model we will utilize to represent the clean system is:

$$H = H_{Kin} + H_{SOC} + H_B \tag{1}$$

with

$$H_{Kin}(\boldsymbol{k}) = 2t \begin{pmatrix} 0 & \cos \boldsymbol{k} \cdot \boldsymbol{a_1} & \cos \boldsymbol{k} \cdot \boldsymbol{a_2} \\ \cos \boldsymbol{k} \cdot \boldsymbol{a_1} & 0 & \cos \boldsymbol{k} \cdot \boldsymbol{a_3} \\ \cos \boldsymbol{k} \cdot \boldsymbol{a_2} & \cos \boldsymbol{k} \cdot \boldsymbol{a_3} & 0 \end{pmatrix} \tag{2}$$

$$H_{SOC}(\boldsymbol{k}) = 2\lambda i \begin{pmatrix} 0 & \cos \boldsymbol{k} \cdot (\boldsymbol{a_2} + \boldsymbol{a_3}) & -\cos \boldsymbol{k} \cdot (\boldsymbol{a_3} - \boldsymbol{a_1}) \\ -\cos \boldsymbol{k} \cdot (\boldsymbol{a_2} + \boldsymbol{a_3}) & 0 & \cos \boldsymbol{k} \cdot (\boldsymbol{a_1} + \boldsymbol{a_2}) \\ \cos \boldsymbol{k} \cdot (\boldsymbol{a_3} - \boldsymbol{a_1}) & -\cos \boldsymbol{k} \cdot (\boldsymbol{a_1} + \boldsymbol{a_2}) & 0 \end{pmatrix} \tag{3}$$

$$H_B(k) = B \begin{pmatrix} 1 & 0 & 0 \\ 0 & 1 & 0 \\ 0 & 0 & 1 \end{pmatrix} \otimes \sigma_z \tag{4}$$

where the 3×3 structure is due to the Co-sites within the unit cell, $\sigma_i$ are Pauli-matrices within the spin-space of the states.

Since the overall Hamiltonian is diagonal in spin space, we can and will consider each spin separately. The free Green function matrix can then be shown to be:

$$\underline{\underline{G_0^\sigma}}(k,\omega) = \Sigma_\nu \frac{\widehat{P_\nu}(k)}{\omega - E_\nu^\sigma(k) + i\eta} = [\omega - H^\sigma(k) + i\eta]^{-1} \tag{5}$$

where $\widehat{P_\nu}$ is the projection operator onto the Hamiltonian eigenstate $\nu$, and the real space Green function is:

$$\underline{\underline{G_0^\sigma}}(\boldsymbol{k},\omega) = \Sigma_\nu \frac{\widehat{P_\nu}(\boldsymbol{k})}{\omega - E_\nu^\sigma(\boldsymbol{k}) + i\eta} = [\omega - H^\sigma(\boldsymbol{k}) + i\eta]^{-1} \tag{6}$$

where $\hat{P}_\nu$ is the projection operator onto the Hamiltonian eigenstate $\nu$, and the real space Green function is:

$$\underline{\underline{G_0^\sigma}}(r_i - r_j, \omega) = \frac{1}{\nu}\sum_\nu [\omega - H^\sigma(k) + i\eta]^{-1} \exp[i\mathbf{k} \cdot (r_i - r_j)] \tag{7}$$

We are interested in the full, on-site Green function in the presence of an impurity at $\mathbf{r} = (0,0)$. The Dyson equation of the general, real-space Green function is:

$$\underline{\underline{G^\sigma}}(r_i, r_j, \omega) = \underline{\underline{G_0^\sigma}}(r_i - r_j, \omega) + \underline{\underline{G_0^\sigma}}(r_i, \omega) + \underline{\underline{V}}\,\underline{\underline{G^\sigma}}(0, r_j, \omega) \tag{8}$$

where $\underline{\underline{V}}$ is the contribution to the Hamiltonian from the impurity, which we assume to be nonmagnetic ($V_{mag} = 0$ and only considering the potential scattering $V_{Pot}$):

$$\underline{\underline{V}} = V_{Pot} \begin{pmatrix} 1 & 0 & 0 \\ 0 & 1 & 0 \\ 0 & 0 & 1 \end{pmatrix} \otimes \sigma_0 + V_{Mag} \begin{pmatrix} 1 & 0 & 0 \\ 0 & 1 & 0 \\ 0 & 0 & 1 \end{pmatrix} \otimes \sigma_z \tag{9}$$

where we included the magnetic term for generality. Crucially now:

$$\underline{\underline{G^\sigma}}(0, r_j, \omega) = \underline{\underline{G_0^\sigma}}(-r_j, \omega) + \underline{\underline{G_0^\sigma}}(0, \omega)\underline{\underline{V}}\,\underline{\underline{G^\sigma}}(0, r_j, \omega) = \underline{\underline{G_0^\sigma}}(r_j, \omega) + \underline{\underline{G_0^\sigma}}(0, \omega)\underline{\underline{V}}\,\underline{\underline{G^\sigma}}(0, r_j, \omega)$$

$$\Rightarrow \underline{\underline{G^\sigma}}(0, r_j, \omega) = [1 - \underline{\underline{G_0^\sigma}}(0, \omega)\underline{\underline{V}}]^{-1}\underline{\underline{G_0^\sigma}}(r_j, \omega) \tag{10}$$

and thus

$$\underline{\underline{G^\sigma}}(r_i, r_j, \omega) = \underline{\underline{G_0^\sigma}}(r_i - r_j, \omega) + \underline{\underline{G_0^\sigma}}(r_i, \omega)\underline{\underline{V}}[1 - \underline{\underline{G_0^\sigma}}(0, \omega)\underline{\underline{V}}]^{-1}\underline{\underline{G_0^\sigma}}(r_j, \omega)$$

$$\equiv \underline{\underline{G_0^\sigma}}(r_i - r_j, \omega) + \underline{\underline{G_0^\sigma}}(r_i, \omega)\underline{\underline{T^\sigma}}(\omega)\underline{\underline{G_0^\sigma}}(r_i, \omega) \tag{11}$$

From this we finally have an expression for the on-site Green function:

$$\underline{\underline{G^\sigma}}(r_i, r_j, \omega) = \underline{\underline{G_0^\sigma}}(0, \omega) + \underline{\underline{G_0^\sigma}}(r_i, \omega)\underline{\underline{T^\sigma}}(\omega)\underline{\underline{G_0^\sigma}}(r_i, \omega) \tag{12}$$

At this point, we must calculate the T-matrix and the impurity-free Green function, which we will do numerically.

We set the lattice constant $a$ of the kagome lattice to be 1, while the hopping integral, $t = 1$. Thus length is measured in $a$, while energy is measured in $t$. Figure 4a shows the dispersion along the K−Γ−K direction of the Brillouin zone calculated directly from diagonalizing $H(\mathbf{k})$, which captures the key band features in the first-principle calculation in Ref. 11. We now turn to the DOS from the full Green function. Figure 4b shows the spin-resolved LDOS calculated from the impurity-free Green function and the full Green function at the impurity site, $r = (0,0)$, for an impurity potential of $V_{Pot} = 5$. A spin-down DOS peak is found in the band gap between spin up and spin down. The spin-up resonance may be damped by coupling with other bands, and more realistic band structures need to be considered to fully produce the experimental results, which deserves further model investigation. This minimum model calculation supports that the nonmagnetic scattering in this magnetic kagome lattice system can generate strong magnetic resonance.

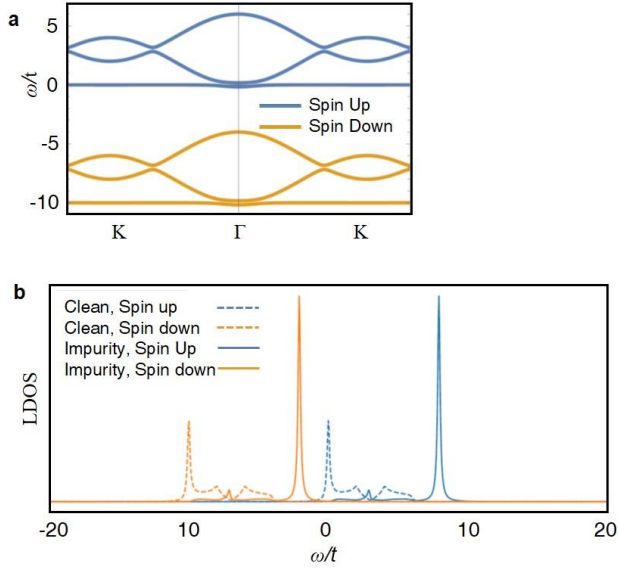

**Supplementary Figure 4. Nonmagnetic scattering induced magnetic resonance. a,** Calculated band structure along the high symmetry direction with parameters: $t = 1$, $\mu = 3$, $\lambda = 0.05$, and $B = 5$. **b,** The spin-resolved LDOS in the clean case and at the impurity site with $V_{Pot} = 5$.

**Supplementary Note 3**

**Fundamental tight-binding model for impurity state coupling.** We consider a tight-binding model and construct an effective Hamiltonian for coupled impurities:

$$H = H_k + H_{SOC} + H_Z \tag{13}$$

Here $H_k$ represents the hopping term:

$$H_k = \sum_{ij} t c_i^\dagger c_j \tag{14}$$

where $t$ is inter-impurity hopping amplitude and $c_i(c_j^\dagger)$ is the electron annihilation(creation) operator in the spinor notation at impurity site i(j). $H_{SOC}$ is the spin-orbit interaction:

$$H_{SOC} = i \sum_{ij} \lambda v_{ij} \left( c_i^\dagger \sigma_z c_j \right) \tag{15}$$

where $\lambda$ is spin-orbit coupling amplitude, $\sigma_z$ is the spin Pauli matrix and $v_{ij} = (\boldsymbol{d_i} \times \boldsymbol{d_j}) \cdot \boldsymbol{z}$ with $\boldsymbol{d_i}$ and $\boldsymbol{d_j}$ denoting the unit vectors along the two bonds that the electron traverses from site i to site j. $H_Z$ is the Zeeman term induced by out-of-plane magnetization of the material:

$$H_Z = \sum_i B c_i^\dagger \sigma_z c_i \tag{16}$$

where $B$ is the effective Zeeman field. We first consider double impurity case, the Hamiltonian reads:

$$H_{double} = \begin{pmatrix} 0 & t \\ t & 0 \end{pmatrix} \otimes I + BI \otimes \sigma_z \tag{17}$$

The spin-orbit coupling does not play a role in the double impurity case. We assume the effective Zeeman field polarizes the impurity bound state such that two spin channels are well separated from each other. For each spin channel, we obtain a bonding state with energy $t$ and an anti-bonding state with energy $-t$.

For the triple impurity case, the spin-orbit coupling effect has to be considered. The Hamiltonian in real space is:

$$H = \begin{pmatrix} 0 & t & t \\ t & 0 & t \\ t & t & 0 \end{pmatrix} \otimes I + \begin{pmatrix} 0 & i\lambda & -i\lambda \\ -i\lambda & 0 & i\lambda \\ i\lambda & -i\lambda & 0 \end{pmatrix} \otimes \sigma_z + BI \otimes \sigma_z \tag{18}$$

Considering only one spin channel, we get three energy levels: $2t, -t \pm \sqrt{3}\lambda$. Without spin-orbit coupling, the system has one single state with energy $2t$ and two degenerate eigenstates with energy $-t$. This degeneracy is protected by $C_{3v}$ symmetry. The spin-orbit coupling breaks the mirror symmetry and lifts the two degenerate levels with energy difference $2\sqrt{3}\lambda$.

From this basic coupled-impurity model, the bound state energies for single, doubly, and triply coupled impurities will be (0), ($t$, -$t$), ($2t$, -$t$-$\sqrt{3}\lambda$, -$t$+$\sqrt{3}\lambda$), respectively. Additional local hole doping effects need to be considered to fully explain the observed states (A), (B, C), and (D, E, F) in Fig. 3b, which could be related to the fact that In is a hole dopant in Sn layer[20-22]. Remarkably, the key ratio (D-(E+F)/2)/(B-C) = (310-(220+170)/2)/(315-240) = 1.5 ± 0.1, which is the expected value *3t/2t*=1.5, confirming the internal consistency of our experiment with respect to this basic coupled impurity model.